\documentclass[12pt,preprint,flushrt]{aastex}

\begin{document}
\newcommand{\etal}{{\it et al.}}
\include{idl}

\title{Evidence for a Millisecond Pulsar in 4U 1636-53 During a Superburst}
\author{Tod E. Strohmayer, and Craig. B. Markwardt$^1$}
\affil{Laboratory for High Energy Astrophysics, NASA's Goddard Space Flight 
Center, Greenbelt, MD 20771; stroh@clarence.gsfc.nasa.gov, 
craigm@milkyway.gsfc.nasa.gov \\ $^1$also, Dept. of Astronomy, 
University of Maryland, College Park, MD 20742}

\begin{abstract}

We report the discovery with the Proportional Counter Array on board the Rossi
X-ray Timing Explorer of highly coherent 582 Hz pulsations during the 
February 22, 2001 (UT) ``superburst'' from 4U 1636-53. The pulsations are 
detected during an $\approx$ 800 s interval spanning the flux maximum of the 
burst. Within this interval the barycentric oscillation frequency increases 
in a monotonic fashion from 581.89 to 581.93 Hz. The predicted orbital 
motion of the neutron star during this interval is consistent with such an 
increase as long as optical maximum corresponds roughly with superior 
conjunction of V801 Arae, the optical companion to the neutron star in 
4U 1636-53. We show that a range of circular orbits with velocity $90 < v_{ns} 
\sin i < 175 $ km s$^{-1}$ and fractional phase $0.336 > \phi_0 > 0.277$ for 
the neutron star can provide an excellent description of the frequency and 
phase evolution. The brevity of the observed pulse train with respect to the 
3.8 hour orbital period unfortunately does not allow for more precise 
constraints. The average pulse profile is sinusoidal and the time averaged 
pulsation amplitude, as inferred from the half amplitude of the sinusoid is 
$1 \%$, smaller than typical for burst oscillations observed in normal 
thermonuclear bursts.  We do not detect any higher harmonics nor the putative 
subharmonic near 290 Hz. The $90\%$ upper limits on signal amplitude at the 
subharmonic and first harmonic are $0.1\%$ and $0.06\%$, respectively. 
The highly coherent pulsation, with a $Q \equiv \nu_0 / \Delta\nu > 4.5 
\times 10^5$ provides compelling evidence for a rapidly rotating neutron star 
in 4U 1636-53, and further supports the connection of burst oscillation 
frequencies with the spin frequencies of neutron stars. Our results provide 
further evidence that some millisecond pulsars are spun up via accretion in 
LMXBs. We also discuss the implications of our orbital velocity constraint 
for the masses of the components of 4U 1636-53.

\end{abstract}


\section{Introduction}

The low mass X-ray binary (LMXB) 4U 1636-53 is one of approximately 10 neutron 
star X-ray binaries which have revealed high frequency pulsations during 
thermonuclear bursts, hereafter ``burst oscillations.'' The $581$ Hz burst
oscillations in 4U 1636-53 were discovered with the {\it Rossi X-ray Timing 
Explorer} (RXTE) by Zhang et al. (1997). These oscillations are almost 
certainly caused by rotational modulation of a nonuniform X-ray emission
pattern on the neutron star surface. Near the onset of bursts this 
pattern likely takes the form of a hot spot (or possibly a pair of hot 
spots) induced on the neutron star surface by the ignition of nuclear burning
(see Strohmayer, Zhang \& Swank 1997). In the cooling part of
the burst lightcurve the modulation pattern is less certain, but it may be
induced by $r$-modes excited by the thermonuclear flash (see Heyl 2002), or 
perhaps ``extreme weather,'' in the form of hydrodynamic vortices induced by 
thermonuclear heating (see Spitkovsky, Levin \& Ushomirsky 2001). 
In particular, the large modulation amplitudes, high coherence and long term 
stability of the frequency are consistent with the rotation scenario (see 
Strohmayer et al. 1998a; Strohmayer \& Markwardt 1999; Muno et al. 2000 and 
Giles et al. 2002). 

There are now three known accreting millisecond pulsars; SAX J1808.4-3658 
(Wijnands \& van der Klis 1998; Chakrabarty \& Morgan 1998), XTE J1751-305 
(Markwardt \& Swank 2002), and XTE J0929-314 (Remillard, Swank \& Strohmayer 
2002). The latter two having been discovered only very recently. 
Their spin periods are 401 Hz, 435 Hz and 185 Hz, respectively. 
Several thermonuclear bursts have been observed from SAX J1808.4-3658 with the 
BeppoSAX Wide Field Cameras (in 't Zand et al. 1998). In the brightest of the 
three bursts seen from SAX J1808.4-3658 in 't Zand et al. (2001) found evidence
for the presence of 401 Hz burst oscillations, providing additional support to 
the idea that burst oscillation frequencies are set by the spin frequencies 
of neutron stars. 

Strohmayer et al. (1998b) suggested that observations of burst oscillations
in bursts at different binary orbital phases might allow a determination of 
neutron star orbital velocities by measuring the orbital Doppler shifts in the 
burst oscillation frequency. Recently, Giles et al. (2002) have studied the
distribution of the highest observed, or ``asymptotic,'' oscillation 
frequencies in a large sample of bursts from 4U 1636-53. Their sample spans a 
time baseline of more than four years. They find a high degree of frequency 
stability over this timespan and attempt to constrain the neutron star orbital 
velocity using a portion of the observed asymptotic frequency distribution. 

Over the last few years a new regime of nuclear burning on neutron stars has
been revealed with the observation of very long (3 - 5 hr) thermonuclear 
bursts from six LMXBs (see Cornelisse et al. 2000; Strohmayer \& Brown 2002;
Wijnands 2001; Kuulkers et al. 2002; Cornelisse et al. 2002). These bursts 
likely result from unstable burning of the ashes (as for example, carbon)
of hydrogen and/or helium burning (see Cumming \& Bildsten 2001; Strohmayer \& 
Brown 2001; Kuulkers 2001). Of the six superburst sources so far only 
4U 1636-53 has produced more than one event. 
Strohmayer (see Strohmayer \& Brown 2001) and Wijnands
(2001) independently identified the superburst from 4U 1636-53 which ocurred on
February 22, 2001; the former in pointed RXTE observations, the latter in the 
RXTE/ASM data. Using the ASM data Wijnands (2001) also identified a similar 
event from 4U 1636-53 which occurred 4.7 years earlier, thus providing
the first constraint on the recurrence time for superbursts. Only two 
superbursts have been observed in detail with large area, high throughput 
instrumentation; a superburst from 4U 1820-30 (Strohmayer \& Brown 2002), and
the February 22, 2001 event from 4U 1636-53. Both were observed with the 
Proportional Counter Array (PCA) on board RXTE. 

Here we report the discovery of
coherent pulsations at $582$ Hz during the February 22, 2001 superburst 
from 4U 1636-53. The plan of this paper is as follows. In \S 1 we briefly
describe the general properties of the superburst and the data available for
high time resolution studies. We then discuss our discovery of the $582$ Hz
pulsations and describe in detail the time evolution of the pulsation 
frequency and the expected behavior given the known binary orbital ephemeris 
for 4U 1636-53. In \S 3 we describe our phase coherent timing study and show 
that orbital modulation of the frequency fits the data extremely well and that 
the pulsation is coherent. We close in \S 4 with a summary and discussion of 
the implications of our findings. A detailed study of the spectral evolution
and energetics of this superburst will be presented in a future publication.

\section{The February 22, 2001 Superburst}

As part of an approved observing program for 4U 1636-53, RXTE observations
were being conducted on February 22, 2001 when a several hour long superburst
was observed. During these observations data were sampled in two high time 
resolution modes, both with a time resolution of 1/8192 seconds. 
The first was an event mode (E\_125us\_64M\_0\_1s) with 64 energy channels and 
the second a ``burst catcher'' mode (CB\_125us\_1M\_0\_249\_H) designed to 
capture high time resolution lightcurves of bursts across the entire PCA 
bandpass. Due to the duration and intensity of the 
superburst, the telemetry rate was very high and portions of RXTE's onboard 
data recorder, which operates as a circular buffer, were filled before they 
could be downloaded. Unfortunately, this resulted in the loss of some high 
time resolution data during the superburst. A similar circumstance ocurred when
RXTE observed a superburst from 4U 1820-30 (see Strohmayer \& Brown 2002). 
In spite of these difficulties, two intervals of high time resolution data 
were obtained; the first beginning just before the peak of the superburst, 
and the second in the decaying tail. Data covering the entire burst were 
obtained in the PCA Standard modes. Standard1 mode data provides 1/8 second 
resolution lightcurves across the entire PCA bandpass, while Standard2 mode 
provides 129 channel spectra every 16 seconds. These modes use much less 
telemetry capacity than the high time resolution modes, and are written to a 
different virtual channel in the data recorder which was not overwritten. 

The 2-60 keV lightcurve of the superburst from Standard1 data is 
shown in Figure 1a with 1 second time resolution. The rise in count
rate beginning near 5500 seconds in the figure represents the source coming out
of Earth occultation. For several minutes after this the intensity level was 
about 4,800 counts/sec, significantly higher than the persistent source rate in
the previous orbit. This difference is much too high to be accounted for by 
variations in the PCA background rate and suggests that the burst may have 
already begun within the Earth occultation gap. The sharp increase in count 
rate which follows is real and can be seen with 1/8 second resolution in 
Figure 1b. This ``precursor'' began at 16:52:12 February 22, 2001 (UTC). 
The precursor has a timescale characteristic of normal (10 - 20 s duration) 
type I bursts from 4U 1636-53, except that it shows an interesting double 
peaked structure. This behavior appears similar to that seen in the superburst 
from 4U 1820-30, which also showed a normal Type I burst precursor at its 
onset (see Figure 2 in Strohmayer \& Brown 2002). Further study of these 
interesting aspects of the superburst will be described in a future
publication. 

\subsection{Detection of 582 Hz Pulsations}

The first high time resolution data interval began at $17:00:45$ UTC on
February 22, 2001 and lasted $\approx 2354$ seconds. This interval, which 
spans the peak of the superburst, is marked by the first set of vertical 
dashed lines in Figure 1a. We began our timing study 
by computing FFT power spectra of this interval. We used 1024 second 
intervals of the full 2 - 60 keV bandpass event mode data to compute two 
power spectra, each with a Nyquist frequency of 4096 Hz. We then searched these
power spectra for significant signals in the vicinity of 581 Hz.  It was clear
almost immediately that the power spectrum of the first 1024 s interval 
contained a significant signal at $\approx 582$ Hz.  The power
spectrum of this interval, rebinned in frequency space by a factor of 8, is
shown in Figure 2a and reveals a highly significant peak near 582 Hz. Figure 2b
shows an expanded view of the region around 582 Hz, revealing a pair of peaks 
separated by $\approx 0.03$ Hz. The highest peak in this power spectrum has a 
single trial chance probability of $\approx 1 \times 10^{-21}$, so there is no 
doubt about the detection. The implied frequency width of $0.03$ Hz 
indicates a quality factor, $Q \equiv \nu_0 / \Delta\nu > 1.94\times 10^4$, 
which without any additional timing analysis, is about a factor of 5 larger 
than that seen to date in burst oscillations from normal Type I bursts 
(see Strohmayer \& Markwardt 1999; Muno et al. 2000). We did not detect the
signal in the remainder of the first high time resolution interval. The 
upper limit to the rms pulsed amplitude was $0.34\%$. The 2nd high time
resolution interval is marked by the 2nd pair of vertical dashed lines in 
Figure 1a. We searched this interval but detected no signal similar to that
found in the first interval. The upper limit to the rms pulsed amplitude in 
this interval was $0.43\%$. 

\subsection{Time Evolution of the Pulsation Frequency}

In order to investigate the time dependence of the pulsation frequency and to
determine whether it persists throughout the entire interval we
computed a dynamic power spectrum using the $Z_1^2$ power statistic (see 
Buccheri 1983; Strohmayer \& Markwardt 1999). However, in anticipation of 
the need for a precise timing analysis, we first barycentered the data using
the standard RXTE software (faxbary) and the JPL DE200 solar system ephemeris 
(Standish et al. 1992).
This analysis revealed that the two peaks in Figure 2 result from two segments
of detectable pulsations during each of which the pulsation frequency is
observed to increase in a roughly linear fashion. The $800$ s interval 
which contains these two pulse trains is marked by the vertical dotted lines in
Figure 1a. Figure 3 shows a contour map of the dynamic $Z_1^2$ (see below) 
spectrum as well as the PCA countrate as a function of time. To produce Figure 
3 we used 64 s intervals to compute $Z_1^2$ and we started a new interval every
16 s. This figure shows that the pulsations persist for {\it hundreds} of 
seconds, much longer than the $\sim 10$ s pulse trains observed in normal 
bursts. Moreover, the overall frequency drift spans only about $0.04$ Hz, 
which is very small compared with the few Hz drifts observed in normal 
burst oscillations (Strohmayer \& Markwardt 1999; Muno et al. 2000). 
The magnitude of this drift is similar to what could be produced solely by the 
Doppler motion of the neutron star in its binary orbit. 
A simple estimate, just using the observed magnitude of the frequency drift, 
would require $v_{ns}\sin i / c = \Delta\nu / \nu = 6.9 \times 10^{-5}$, which 
is equivalent to a projected velocity of $21$ km s$^{-1}$. This number 
underestimates the velocity, since we have ignored the rate at which the 
frequency changes. Nevertheless, it suggests that we may be seeing a coherent 
pulsation modulated by orbital motion of the neutron star. 

\subsection{Expected Frequency Evolution Based on Binary Orbital Ephemeris}

The 3.8 hr orbital period of V801 Arae, the optical companion to the 
neutron star in 4U 1636-53, is well known from optical observations. Recently, 
Giles et al. (2002) have used new optical observations to test and slightly 
modify the ephemeris determined from all historical optical observations by 
Augusteijn et al. (1998).  Their resulting ephemeris: HJD = 2446667.3179(33) 
$\pm (N \times 0.15804693(16) )$, predicts the epoch of maximum optical 
brightness in units of heliocentric Julian day.  Although the formal 
statistical error in projecting this ephemeris forward to the time of the 
superburst is of order $4\times 10^{-4}$, there is an $\approx 0.05$ systematic
fractional phase uncertainty which results from the fact that the optical 
lightcurve does not have any sharp fiducial features and also has often shown 
night to night variations (see Giles et al. 2002 for a discussion). 

We used this ephemeris to predict the orbital phase of the $800$ s 
interval during the superburst which shows pulsations. Hereafter all phases
will be given as fractional phases (ie. ranging from 0 to 1). The interval 
begins at MJD 51962.710160410, which corresponds to a fractional phase of 
$0.352$ relative to optical maximum (phase zero is optical maximum). 
One addtional piece of information is required to predict the corresponding 
frequency evolution, the relationship between the optical and dynamical 
ephemerides. For systems like 4U 1636-53, which show sinusoidal optical 
modulations and which have inclinations $\lesssim 60^{\circ}$ the standard 
interpretation has been that optical maximum corresponds to superior 
conjunction of the optical secondary, that is, when the secondary is furthest 
from the observer.  Figure 4 shows the predicted frequency behavior versus 
orbital phase of a coherent pulsation associated with the neutron star (as, 
for example, its spin frequency) given this relationship between the optical 
and dynamical phases. The predicted phase range of the superburst pulsation 
interval is denoted with vertical solid lines. The predicted frequency 
evolution is strikingly similar to that observed, moreover, even an uncertainty
in the phase of $\approx 0.1$ would not change the qualitative conclusion that 
the frequency {\it should} be increasing because of the orbital motion. The 
original Augusteijn (1998) ephemeris predicts an orbital phase of 0.26 for the 
start of the oscillation interval. This is still in the range of orbital 
phases which would predict an increasing pulsation frequency. These 
considerations strongly suggest that the observed frequency variations are 
consistent with the orbital motion of the neutron star. 

Since frequency drift is a common feature of normal burst
oscillations, it is a fair question to ask whether the frequency
variations observed in this superburst --- which we claim to be
orbitally modulated --- could be mimicked by intrinsic drifts
unrelated to the orbit.  We argue that this possibility is unlikely
for a number of reasons.  First, an intrinsic drift would need to replicate 
the apparent orbital modulation at the phase of the observed burst. As we
show below, the observed frequency variation is well fit with a concave 
upwards curvature which is rarely, if ever, seen in normal burst oscillations 
(see Strohmayer \& Markwardt 1999; Muno et al. 2002). A second consideration 
is that the frequency drift observed in the 4U 1636--536 superburst is much 
smaller than that of all other normal burst oscillations.  Examining bursts 
from many different X-ray binaries, Muno et al. (2002) found 68 with 
oscillations, and measured the frequency drifts.  Of those, about 30 had 
oscillations which began more than one second from the burst onset.  
However, {\it none} had
drifts smaller than $\sim$0.3 Hz.  Therefore we put the chance of
observing a 0.04 Hz drift at 1 in 30, or $\sim$97\% confidence.  To
derive a more robust estimate would require more knowledge of
intrinsic frequency drifts associated with superbursts, however, with
only a single example so far, this is not currently feasible.

The frequency drift in the burst oscillation discussed here is consistent 
with a range of orbital phase, whose size gives a rough estimate of the
likelihood of a chance agreement with orbital modulation.  This
likelihood is related to the accuracy with which the phase can be
predicted based on the optical ephemeris and the length of the pulse
train. Obviously, the more of the orbit covered by the pulse train,
the greater our confidence would be in an orbital modulation
interpretation.  Based on the $0.05 - 0.1$ uncertainty in phase noted
above and the $\approx 900$ s pulse train, we can estimate that only
about a 5\% to 10\% range of orbital phase could produce a signature
consistent with the observed drift.  Below we fit orbital models and
derive a range of phase spanning about 6\% of the orbital phase which
can produce an acceptable fit.  So, assuming all the observed
frequency drift is intrinsic and not orbital there is roughly a 6\%
chance that it could be consistent with the expected orbital
modulation.  In our opinion, however, this should be considered an
upper limit, since it seems quite unlikely that all the observed drift
is intrinsic.

In summary, the observed frequency drift appears to be well consistent
with that expected based on the known orbital parameters of V801
Arae. A chance agreement assuming all the observed drift were
intrinsic to the superburst can be ruled out with confidences at
the 94--97\% level.

\section{Phase Coherent Timing Analysis}

In order to test the orbital modulation conclusion more quantitatively we 
performed coherent timing analyses using the complex $Z_1$ statistic (see 
Strohmayer \& Markwardt 1998 for an example of the use of the $Z_1^2$ statistic
in a similar context), where;
\begin{equation}
Z_1 = X + i Y = \left ( \sum_{j=1}^N \cos \phi_j \right ) + i 
\left ( \sum_{j=1}^N \sin \phi_j \right ) .
\end{equation}
The $Z_1^2$ statistic is defined as $Z_1^2 = Z_1^* Z_1$, and the phase angle is
given by $\psi = \tan^{-1} (Y / X)$. Here $\phi_j = 2\pi \int_0^{t_j} \nu (t') 
dt$, where $\nu (t')$ is the frequency as a function of time, and $t_j$ are 
the observed X-ray event times. 

\subsection{Solution Using Total $Z_1^2$ Power}

We used a circular orbit model to describe the time 
variation of the pulsation frequency and phase, {\it vis};
\begin{equation}
\nu (t) = \nu_0 \left ( 1 - \frac{v_{ns}\sin i}{c}\sin (\omega_{orb}t + \phi_0)
\right ) .
\end{equation}
Here, $\nu_0$, $v_{ns}\sin i / c$, $\phi_0$ and $\omega_{orb}$ are the 
barycentric frequency, the projected neutron star velocity, the orbital phase
at the start of the pulsation interval, and the orbital period, respectively. 
We used this model to predict the pulsation phase at time $t$ and we varied the
parameters in order to maximize $Z_1^2$. We used the data from the $\sim 800$ 
s segment shown in Figure 3, but we excluded events in an $160$ s segment 
between the two pulse trains in which we did not detect significant pulsations.
We fixed the orbital frequency $\omega_{orb}$, since it is known with high 
precision and we cannot hope to constrain it further with such a short data
segment, and we let $\nu_0$, $v_{ns}\sin i / c$, and $\phi_0$ vary. We began by
making a grid search over a range of parameter space which produced, ``by 
eye,'' reasonably plausible solutions. We then used a general function 
extremizer, TNMIN (see Nash 1984; implemented in IDL by C. M.), to optimize the
solutions. We double checked the maximization using a simplex method (AMOEBA)
implemented in FORTRAN. The two methods were in excellent agreement. We find 
that this model can describe the frequency and phase 
evolution very well. Our best fits have $Z_1^2 \approx 276$ which represents 
an enormous improvement compared to a model with no frequency evolution. 

As the data segment is short compared to the 3.8 hour orbital period 
it is possible to find statistically similar solutions with a range of orbital 
parameters. Indeed, there is a broad ridge in $Z_1^2$ space on which the
three parameters are strongly correlated. The best model in the 
sense of having the largest $Z_1^2 = 275$ has $\nu_0 = 582.1454$ Hz, 
$v_{ns}\sin i = 136 $ km s$^{-1}$ and $\phi_{0} = 0.293$. However, models with 
90 km s$^{-1}$ $ < v_{ns}\sin i < $ 175 km s$^{-1}$ are not strongly excluded. 
A look at Figure 4 confirms why this is so. As long as the fitted orbital 
phase range is such that the curvature is concave up, then the other 
parameters, $v_{ns}\sin i / c$, and $\nu_0$ can more or less adjust to 
approximate the observed frequency evolution. 

Given the systematic uncertainty
of order 0.05 in the optical phase and the additional uncertainty of the exact 
relationship between optical and dynamical phase it is not possible to prefer 
{\it a priori} a particular $\phi_{0}$, however, the range above is 
consistent with what is expected based on the optical ephemeris as long as 
optical maximum ocurrs within $\approx 0.1$ of superior conjunction of the 
secondary. These considerations show that the data can be well explained by a
range of reasonable orbits, however, because of the shortness of the pulse 
train we cannot place tighter constraints on the orbital parameters using 
these data alone. Nevertheless the observed pulsations are highly coherent as 
we are able to track $\approx 500,000$ consecutive cycles. 

In Figure 5 we compare the standard FFT power spectrum with the $Z_1^2$ power
spectrum using our best orbital parameters. The pair of weak broad peaks 
near 581.91 Hz are from the standard FFT analysis and are effectively the 
same as the pair of peaks in Figure 2. The narrow peaks near 582.1 Hz come from
the $Z_1^2$ spectrum and result from the coherent addition of the two 
pulse trains. The sidelobe peaks are caused by the need to bridge the 
gap between the two pulse trains. Figure 6 shows the best orbital frequency 
model plotted over the dynamic $Z_1^2$ contours, and demonstrates
visually that it tracks the frequency vs time contours very well. The success 
of the coherent timing analysis provides strong evidence that we are seeing the
orbital modulation of coherent pulsations during the superburst.

\subsection{Phase Evolution Analysis}

We can also investigate the problem in the context of 
phase evolution as a function of time. For a given set of model parameters we 
can compute the phase, $\phi_j$, of each x-ray event. We can then 
break up the full set of events into a set of $M$ bins and compute the phase
$\psi_k$ for each bin. The phases are simply given by;
\begin{equation}
\psi_k = \tan^{-1} \left ( \sum_m \sin\phi_m / \sum_m \cos\phi_m \right )  
\;\; , 
\end{equation}
where the index $m$ runs over all events in bin $k$. We can then compute
$\chi^2 = \sum_{k=1}^M \left ( \psi_k - \psi_{avg} \right )^2 / 
\sigma_{\psi_k}^2$, where $\psi_{avg}$ is the average phase angle computed 
from all $M$ bins. We can then minimize $\chi^2$ to find a best fitting 
model. For a coherent signal the error $\sigma_{\psi_k}$ in the phase angle 
is given simply by $1/\sqrt(Z_1^2)$, a result which we have confirmed with 
monte carlo simulations. This procedure is essentially equivalent to the phase 
connected timing analysis of burst oscillations described by Muno et al. 
(2000). 

To further explore the phase coherence of our orbital solutions we broke up 
our event data into 24 segments and carried out the $\chi^2$ minimization 
described above. We found that the best orbit model was entirely consistent 
with the model obtained from maximizing the $Z_1^2$ statistic. The minimum 
$\chi^2$ was 23.3, which for 20 degrees of freedom (3 parameters and 1 
additional degree of freedom from calculating $\psi_{avg}$) gives a 
statistically acceptable solution. The resulting phase residuals and errors are
shown in Figure 7 in units of milliperiods. 
The rms level of $46$ milliperiods is indicated by the dashed 
horizontal lines. This is equivalent to $79$ $\mu$s. Although the 
residuals are unstructured in the sense they do not show any large systematic 
deviations, there is some weak evidence for phase jitter
(or timing noise in the parlance of radio pulsar timing) with a scale of 
about $20-40$ milliperiods on timescales of $\approx$hundreds of seconds, 
particularly during the first pulse train. Jitter of a similar magnitude 
is also suggested by dynamic power spectral and phase coherent studies of 
some burst oscillations (see for example, Wijnands, Strohmayer \& Franco 2000; 
Muno, Chakrabarty \& Galloway 2002). To find the $90\%$ confidence limits on 
the model parameters we found where $\Delta\chi^2 = 6.25$, which is appropriate
for a 3 parameter fit. This gives a $90\%$ confidence range for each orbit 
parameter of; $582.0390 < \nu_0 \; ({\rm Hz}) < 582.2262$, $90 < v_{ns}\sin i 
\; ({\rm km \; s}^{-1})< 175$, and $0.336 > \phi_0 > 0.277$. 

We also compared the circular orbital model with a strictly linear increase of 
the frequency. The linear model achieves only a minimum $\chi^2 = 58.2$, 
indicating that the circular orbit model performs significantly better. 
Quantifying the significance of the difference in $\chi^2$ with the F-test 
shows that the additional parameter using the circular orbit model is 
significant at the $2.3 \times 10^{-5}$ level. We note that the sense of the 
improvement is such that frequency models with curvature (such that the 
frequency vs time curve is concave up) fit much better than linear models.   

We next used our best timing solution to phase fold the data and obtain
an average pulse profile which we show in Figure 8. The units are counts and 
we repeat 2 cycles for clarity. The pulse profile is consistent with a sinusoid
with a half amplitude of 1\% of the mean number of counts. 
Having computed a coherent solution we also made a sensitive search for 
harmonics and subharmonics, but we did not detect any higher harmonics nor the 
$290$ Hz subharmonic (see Strohmayer \& Markwardt 1999; Miller 1999; 
Strohmayer 2001). The $90\%$ confidence upper limits to the signal amplitude 
at the 290 Hz subharmonic and 1,160 Hz 1st harmonic are $0.1$ and $0.06\%$ of 
the mean count rate (rms), respectively.

\section{Implications and Discussion}

The observation of coherent pulsations during a superburst from 4U 1636-53 
provides additional strong evidence of a rapidly rotating neutron star in 
4U 1636-53. For example, our observation of $\sim 500,000$ pulsation cycles 
at $582$ Hz is roughly analogous to observing a standard $\sim1$ s period 
radio or X-ray pulsar for $\sim 6$ days. There would seem to be little doubt 
that such stability can only reflect the spin period of the neutron star. In 
the remainder of this section we discuss some of the implications of our 
findings. 

\subsection{Implications for Burst Oscillations}

Our coherent timing analysis places a constraint on the barycentric frequency
of the coherent pulsation in 4U 1636-53 of $582.0390 < \nu_0 < 582.2262$ Hz. 
This range is higher than any of the asymptotic burst oscillation frequencies
measured for 4U 1636-53 by Giles et al. (2002). Our $90\%$ upper limit is 
about 0.4 Hz above the highest asymptotic oscillation frequency (see Giles 
et al. 2002), but the limits on the barycentric pulsar frequency are  
reasonably consistent with the {\it total} observed spread around this peak 
(see Figure 6 in Giles et al. 2002). The fact that the implied pulsar 
frequency is within $\approx$1 Hz of all the measured asymptotic frequencies 
provides further evidence that the frequencies observed during the more 
numerous, short duration ($\approx 10$ s) thermonuclear bursts are indeed set 
by the rotational frequency of the neutron star. However, since the 
distribution of {\it all} asymptotic burst oscillation periods from 4U 1636-53 
are not strongly correlated with orbital phase there arguably must be an 
additional source of variation in the asymptotic frequencies of burst 
oscillations above that caused by orbital motion of the neutron star.  

There has been much recent theoretical work to provide an explanation of the
few Hz frequency drifts observed in burst oscillations. For example,
Cumming et al. (2002) have reevaluated their earlier (Cumming \& Bildsten 2000)
theoretical calculations of spin down in the neutron star surface layers 
produced by hydrostatic expansion and angular momentum conservation 
during thermonuclear bursts. They conclude that such expansion alone
is probably not capable of accounting for the largest observed frequency 
drifts.  More recently, Spitkovsky, Levin \& Ushomirsky (2002) have 
investigated the effects of burst heating in the atmosphere and show that this
can drive strong zonal flows which can also produce frequency drifts. They 
suggest that some combination of radial uplift and horizontal flows may 
explain the observed drift. Heyl (2002) has recently proposed that $r$-modes
excited by the thermonuclear burst may be responsible for producing the 
observed modulation, especially in the decaying tails, and that the motions of 
these waves relative to the star can produce the observed frequency drifts.  

Our results provide further impetus for such work.  The discovery of coherent
pulsations argues compellingly for the spin interpretation, which can be 
summarized as follows. The observed modulations are caused by the rotation of 
some anisotropic emission pattern on the neutron star surface. Near burst 
onset the pattern is most likely a localized ``spot'' (or perhaps antipodal 
spots). At late times in the burst tails it may be produced by an oscillation 
mode, perhaps the $r$-modes suggested by Heyl (2002), or the hydrodynamic 
vortices proposed by Spitkovsky, Levin \& Ushomirsky (2002). However, the 
observed frequency drifts and variations in the asymptotic burst 
frequencies must be accomodated within a succesful model. 
It seems most likely that the observed frequency drifts result from the fact 
that the emission pattern is not fixed in the rotating frame of the neutron 
star. Oscillation modes, as for example the $r$-modes, are not fixed in the 
rotating frame, and thermonuclear flame fronts can move across the surface of 
the star. In addition, thermonuclear heating drives both radial and horizontal 
motions, both of which may contribute to the $\lesssim 1\%$ frequency drifts.

\subsection{What Causes the Flux Modulation?}

The coherent pulsations observed during the superburst have an average
amplitude of $1\%$ of the mean 2 - 30 keV countrate, smaller than is typical 
for burst oscillations from normal bursts. This raises the interesting 
question; what causes the flux modulations during the superburst and how does 
it relate to that which operates during normal bursts? 
If a weak coherent modulation were always
present with some fixed amplitude, then if the source intensity increased--as 
during the superburst--the modulation could become detectable. This seems 
unlikely, however, since there are intervals during the burst at roughly 
similar intensities which do not show detectable pulsations.  More likely, 
there is some intrinsic change in the modulation amplitude caused by the 
superburst. There is good evidence that these events result from the release 
of thermonuclear energy at great depth ($\approx 10^{13}$ g cm$^{-2}$, see
Strohmayer \& Brown 2002; Cumming \& Bildsten 2001). If the 
energy release and/or transport is anisotropic then a flux asymmetry at the
surface could be produced. Alternatively, the burst flux might excite an 
oscillation mode (as, for example, suggested by Heyl 2002). In either case, if 
such a process produced a weak $1\%$ modulation which varied in strength 
by about a factor of two, then the intervals of strongest modulation would be 
detectable, but because the spectral power varies as the square of the 
modulation amplitude, a modest drop in the amplitude would render the 
modulation undetectable. Further insight into these issues could result from 
a phase resolved spectral study, which we will pursue in subsequent work.

\subsection{Constraints on the Component Masses}

Our coherent timing analysis places a constraint on the projected neutron 
star velocity. As mentioned above, a range of orbital velocities, initial 
phases and pulsar frequencies are allowed by the data.  Giles et al. (2002)
used the observed distribution of asymptotic burst oscillation frequencies to
constrain the neutron star velocity. They obtained a $99\%$ upper limit on
$v_{ns}\sin i = 50$ km s$^{-1}$ using a subset of 18 bursts which showed a
tight clustering in asymptotic frequency.  Although our inferred velocity 
range is outside their limit, the difference may reflect the exclusion of 
some bursts from the analysis by Giles et al. (2002). Indeed, the overall 
spread in frequency of {\it all} the bursts from 4U 1636-53 is not strongly 
at odds with the higher neutron star velocity inferred from our timing 
analysis here (see for example Figure 6 in Giles et al. 2002).  This does
suggest, however, that it will likely be even more difficult to infer 
reliable neutron star orbital velocities using normal burst oscillation 
frequencies than indicated by the Giles et al. (2002) study.

A measurement of the neutron star orbital velocity has implications for the 
masses of the binary components in 4U 1636-53.  Although the system inclination
is not precisely known, the lack of both eclipses and dips suggests that it
cannot be much greater than $\sim 60^{\circ}$ (see for example, Frank et al.
1987). Figure 9 shows the range of allowable masses for the components, with
the secondary mass along the abscissa and the neutron star mass along the 
ordinate. Velocity contours are shown for the limits calculated from our
orbital timing solutions.  Contours are drawn for two inclinations, 
$70^{\circ}$ (dashed) and $60^{\circ}$ (solid). The allowed masses are
those enclosed by the velocity contours. 

In accreting binaries the mean density of the donor star is constrained if 
the orbital period is known and the star loses mass via Roche lobe overflow 
(see for example, Rappaport, Joss \& Webbink 1982).  The constraint can be
expressed as;
\begin{equation}
\frac{m_2}{M_{\odot}} = 5.427 \; P_{hr}^{-2} \; \left ( \frac{r_2}{R_{\odot}} 
\right )^3 \; ,
\end{equation}
where $P_{hr}$, $m_2$ and $r_2$ are the orbital period in hours, the mass of 
the donor, and the radius of the donor, respectively. If one assumes a 
particular mass - radius relation for the donor star, then the above 
constraint yields a mass estimate. If one takes the canonical mass - radius
relation for main sequence stars (that is, $m_2 \approx r_2$), then for 
4U 1636-53 ($P_{hr} =3.79$) this gives $m_2 = 0.43 M_{\odot}$. If instead one 
uses an empirical mass - radius relation for zero age main sequence (ZAMS)
stars then the mass estimate is reduced somewhat to 0.34 $ <  
( m_2 / M_{\odot} ) < $ 0.37 (see Patterson 1984; Smale \& Mukai 1988). Note,
however, that the empirical mass - radius relation in the range $m_2 < 0.4 
M_{\odot}$ is quite poorly defined (see Patterson 1984; Baraffe et al. 1998).

As can be seen in Figure 9 the mass estimates above are more or less 
consistent with our derived limits, but perhaps uncomfortably so, since
they cluster near the lower velocity limit. 
Kolb, King \& Baraffe (2001) have recently investigated mass estimates in 
compact binaries.  They show that both mass loss and evolution of the donor
prior to the onset of mass exchange can bias the mass estimate. They argue that
typically the real mass will be less than the estimate obtained from the 
Roche lobe constraint and an assumed ZAMS mass - radius relation. These 
effects, if present, would tend to worsen the agreement with our lower range 
of masses in Figure 9.  One can ask what mass - radius relation would yield
a mass estimate more comfortably within the constraints in Figure 9.  
For example, to have a mass of 0.5 $M_{\odot}$ the donor's mass - radius 
relation would need to be $m_2 \approx r_2^{1.15}$. This does not represent a 
large discrepancy from expectations, particularly given the uncertainties at 
the low mass end. 

The constraints shown in Figure 9 are purely statistical in nature. It seems
likely, however, that a systematic uncertainty may also be present because 
the observed pulse train spanned only about $7\%$ of the orbit, and there is 
some indication of phase jitter from our timing analysis. 
These considerations suggests that there could also be a source of systematic 
uncertainty in the derived orbital parameters. Because of this, and the 
sizable uncertainties associated with deriving a mass estimate, for 
example, the assumed mass - radius relation and the orbital inclination, 
we do not think that our present mass constraints are strongly at odds with
theoretical expectations for the nature of the companion star. However, if 
further observations continue to favor a high mass for the companion, 
then it could be an indication that some of the observed frequency drift is 
intrinsic to the superburst and not caused by orbital modulation. 

\pagebreak

\pagebreak
\centerline{\bf Figure Captions}

\figcaption[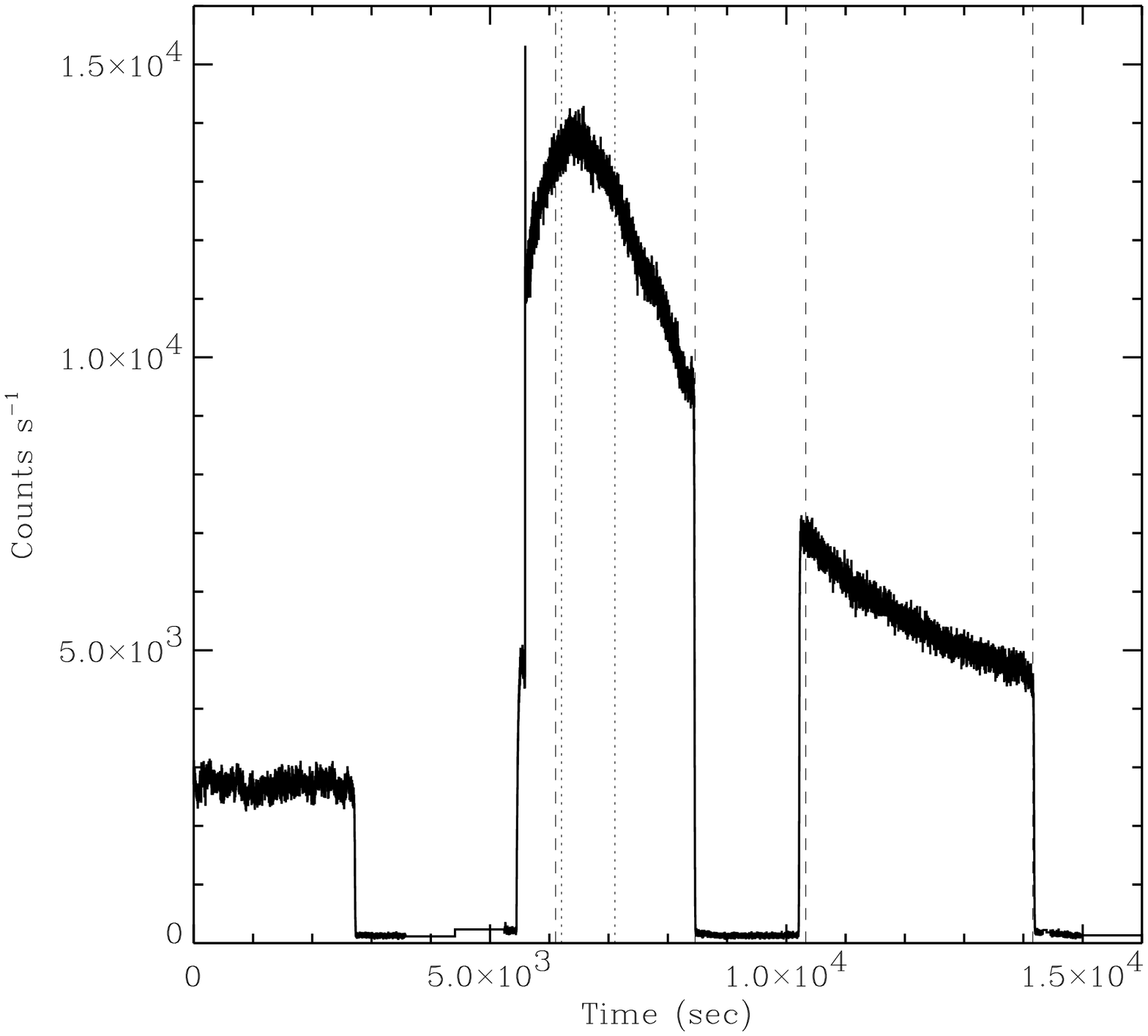]{Lightcurve of the superburst from 4U 1636-53 observed on
February 22, 2001 (top). The data are the 2 - 60 keV PCA countrates from the
Standard1 datamode. In the top panel the time resolution is 1 s. 
The dashed vertical lines denote the intervals with high time resolution event
mode data. The $\sim 800$ s interval in which pulsations 
are detected is shown by the vertical dotted lines. 
An exploded view of the sharp rise near the start of the burst
is also shown (bottom). Here the time resolution is 1/8 s. Note the double 
peaked profile which shows timescales typically seen in normal Type I X-ray 
bursts ($\sim 10$ s). Time is measured from 15:19:00 UTC on Feb. 22, 2001.
 \label{fig1}}

\figcaption[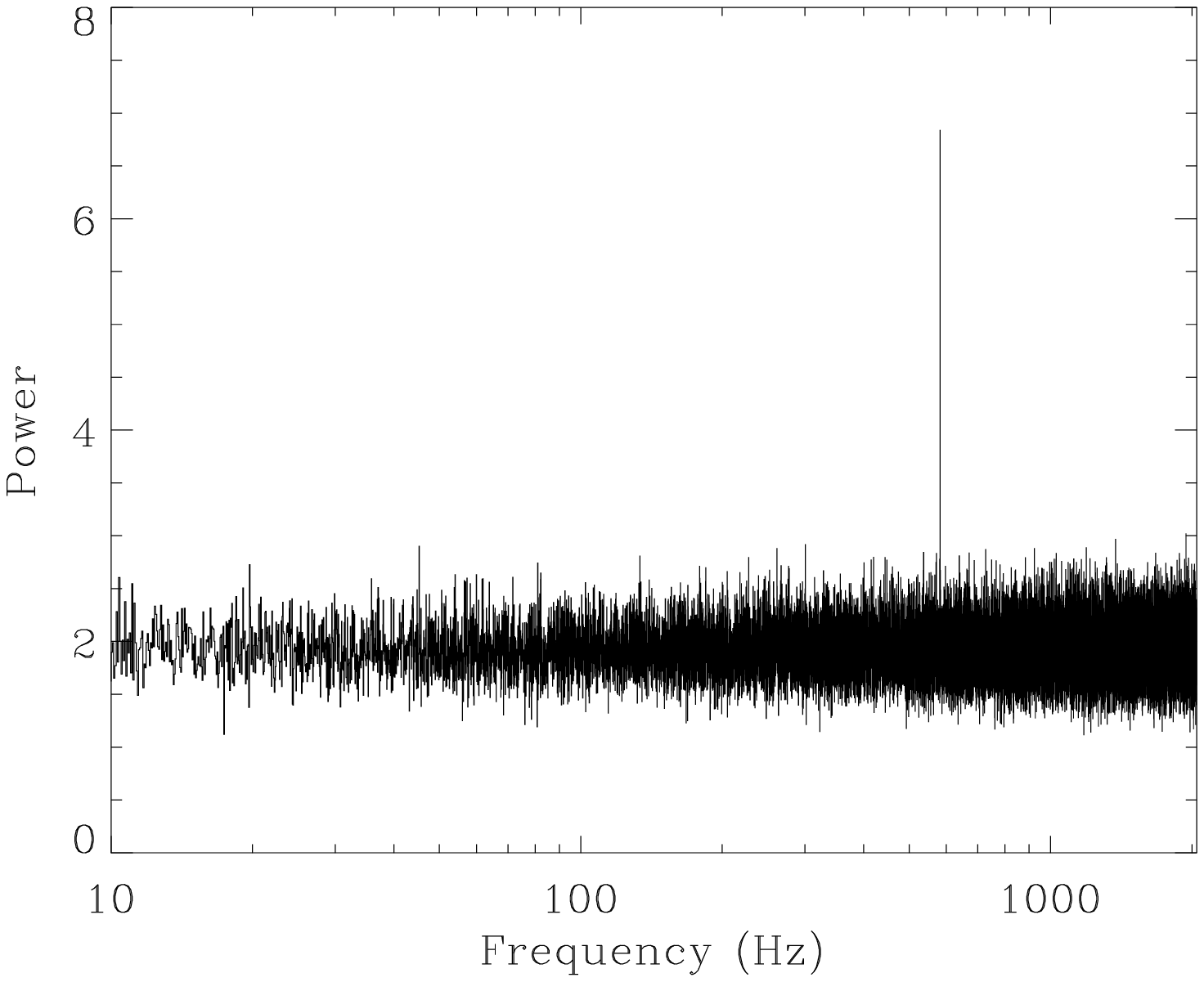]{Power spectrum (top) of the first 1024 second data interval 
starting at 17:00:44 UTC. The spectrum was computed using the full bandpass
event mode data. The Nyquist frequency is 4096 Hz. The frequency resolution 
is 1/128 Hz. The significant peak near 581 Hz is evident. An expanded view 
of the region around 582 Hz is also shown (bottom). The frequency resolution 
in this plot is 1/256 Hz. Two peaks separated by $\sim 0.03$ Hz are clearly 
resolved.
 \label{fig2}}

\figcaption[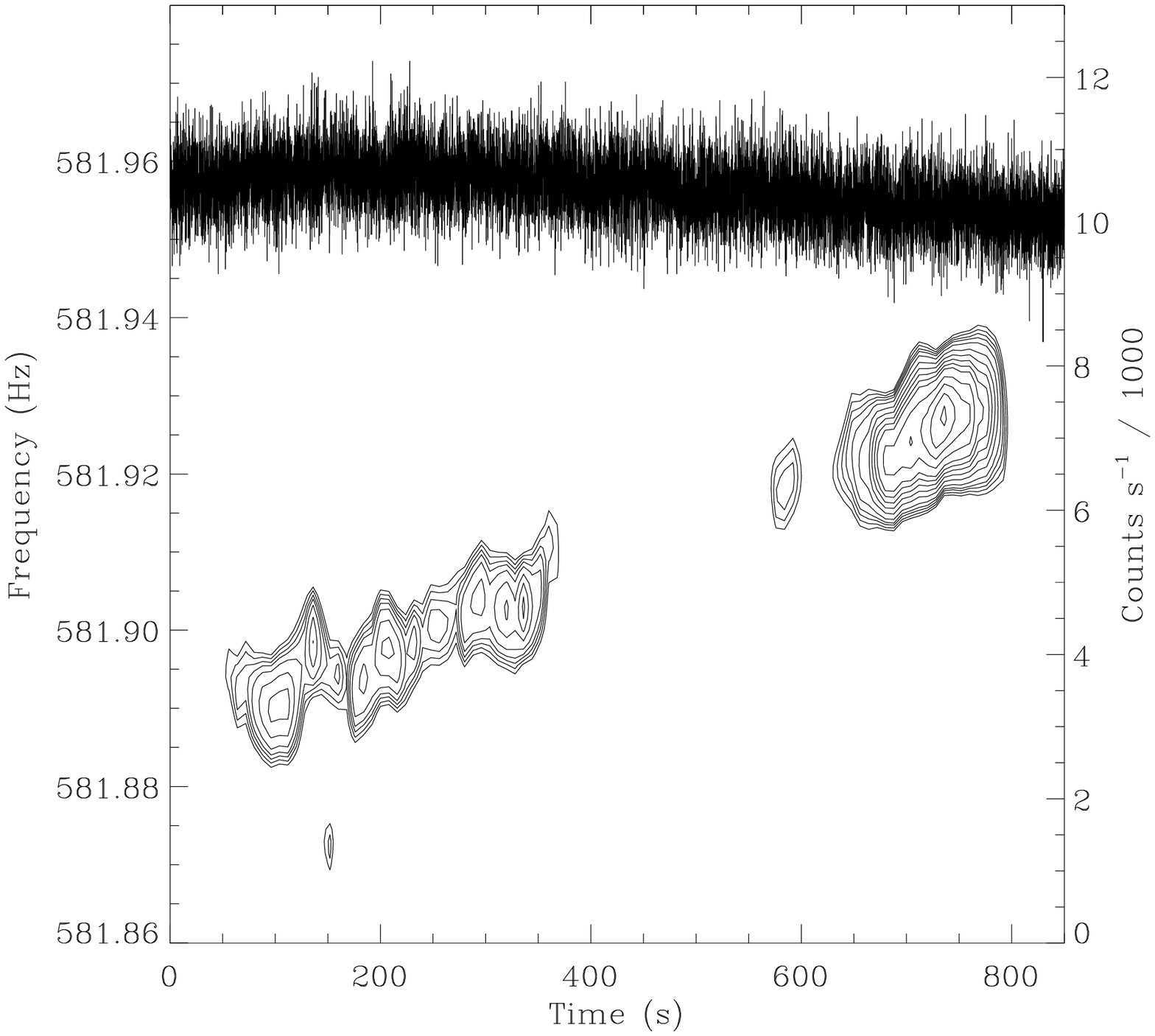]{Dynamic $Z_1^2$ power spectrum of the 800 s interval 
containing significant pulsations. The $Z_1^2$ data were computed using
64 s data intervals and a new interval was started every 16 s. Contours
at 16, 18, 20, 22, 25, 30, 35, 40, 45, 50, 60, 70, 80, and 90 are shown. Two
pulse trains are clearly evident, separated by a gap of $\sim 200$ seconds.
The PCA countrate (4 detectors) is also shown (see right axes). Time is 
measured from $17:02:25$ UTC on Feb. 22, 2001
\label{fig3}}

\figcaption[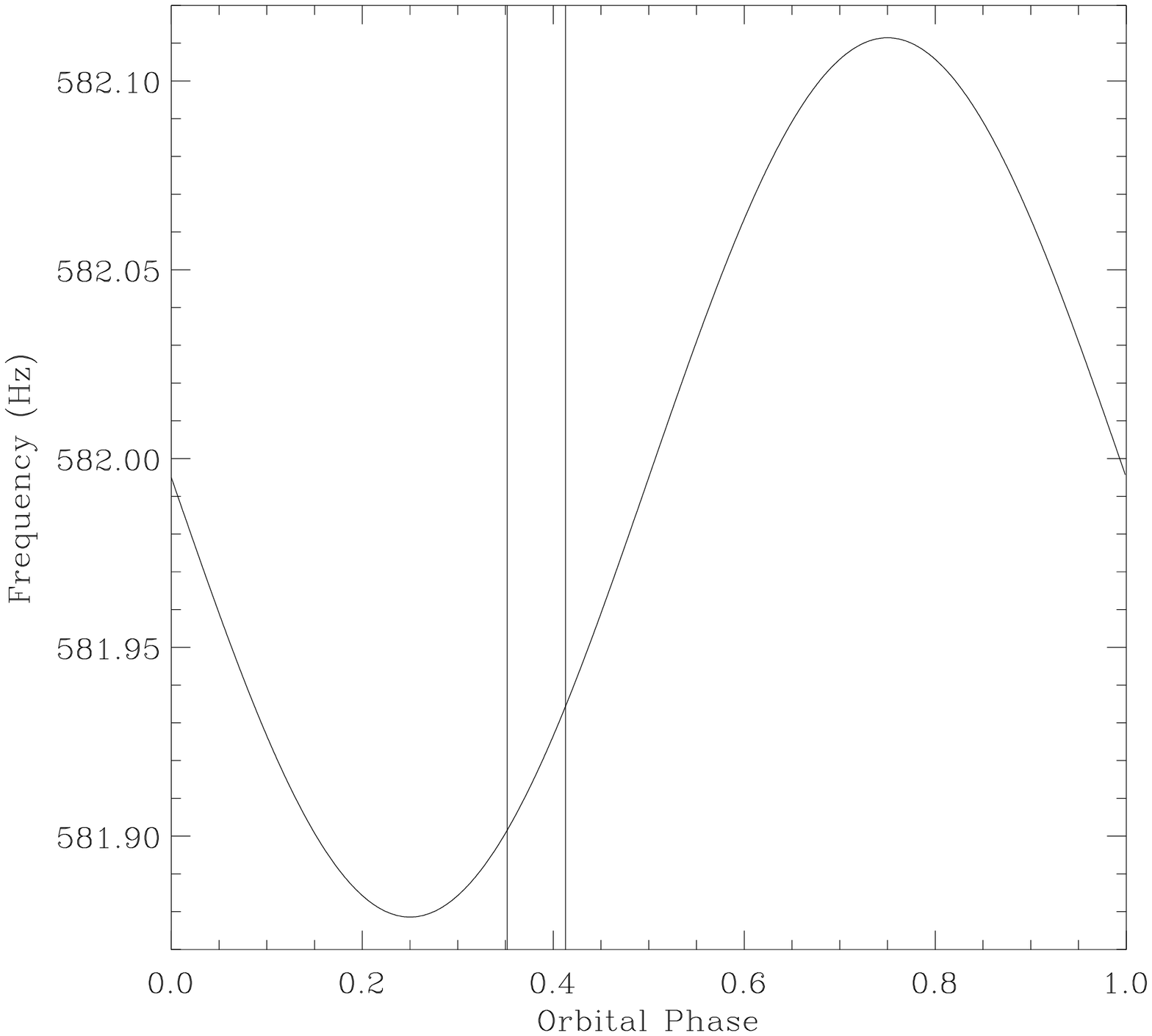]{Predicted orbital frequency evolution of a coherent 
pulsation centered on the neutron star in 4U 1636-53 and the orbital phase 
range of the 800 second pulsation interval during the superburst. We used the
optical ephemeris from Giles et al. (2002) along with the single assumption 
that optical maximum corresponds to superior conjunction of the optical
secondary. The vertical lines denote the phase range of the 
800 s pulsation interval.  The curve shows the frequency modulation which would
be produced with $v\sin i = 62$ km s$^{-1}$ for the neutron star. This is not 
intended to represent a fit to the observed frequency evolution, but 
rather to simply show the qualitative nature of the predicted frequency 
evolution.  Even with an offset of the phase relationship between optical 
maximum and superior conjunction of $\sim 0.1$ the ephemeris predicts that a 
pulsation frequency associated with the neutron star should be increasing.
\label{fig4}}

\figcaption[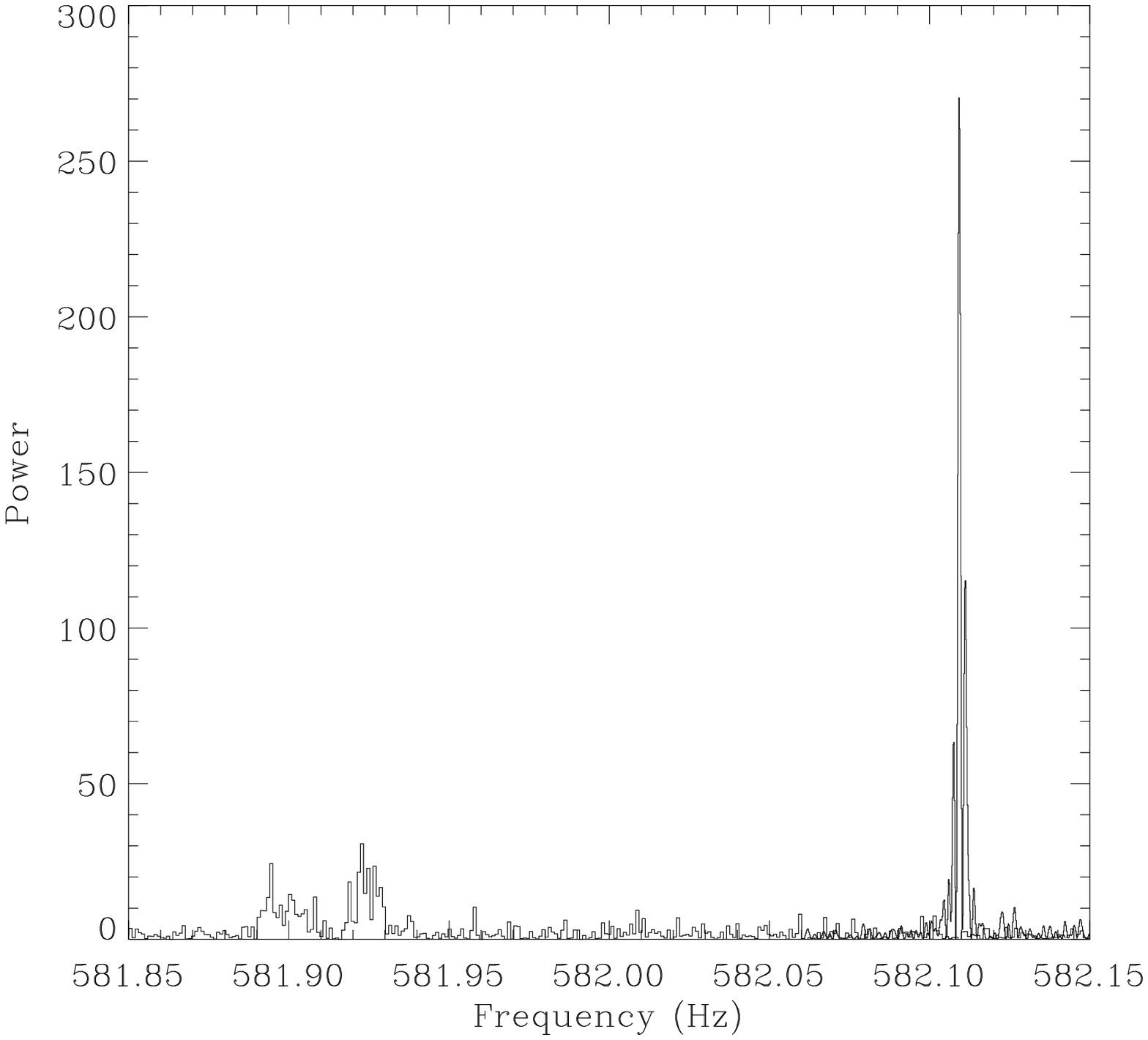]{Comparison of the FFT power spectrum which assumes no
frequency or phase evolution with the $Z_1^2$ power spectrum computed using
the best orbital frequency evolution model. The weak, broad peaks to the
left of the figure are from the standard FFT power spectrum, while the
sharp peaks to the right result from the phase coherent $Z_1^2$ power
spectrum. The sidelobe peaks in the coherent spectrum result from 
the gap between the two pulse trains.  \label{fig5}}

\figcaption[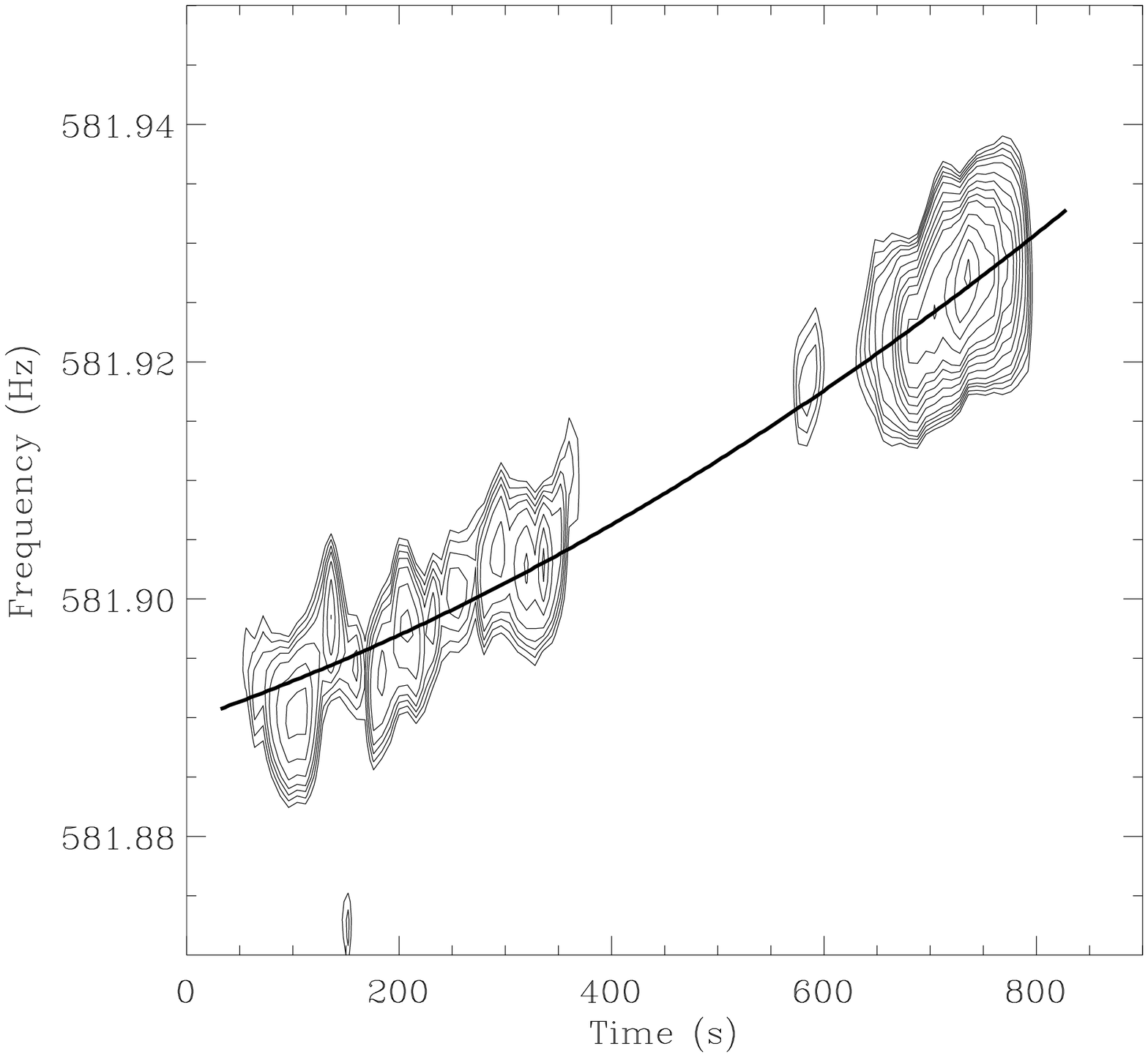]{Dynamic $Z_1^2$ contours as a function of frequency and
time. Also shown is the frequency evolution from the best orbit 
model. The solid curve shows the model with $\nu_0 = 582.1454$ Hz, 
$v_{ns}\sin i = 136$ km s$^{-1}$ and $\phi_{dyn} = 0.293$ and has a 
$Z_1^2 = 275.5$. See the caption to Figure 3 for the reference epoch. 
\label{fig6}}

\figcaption[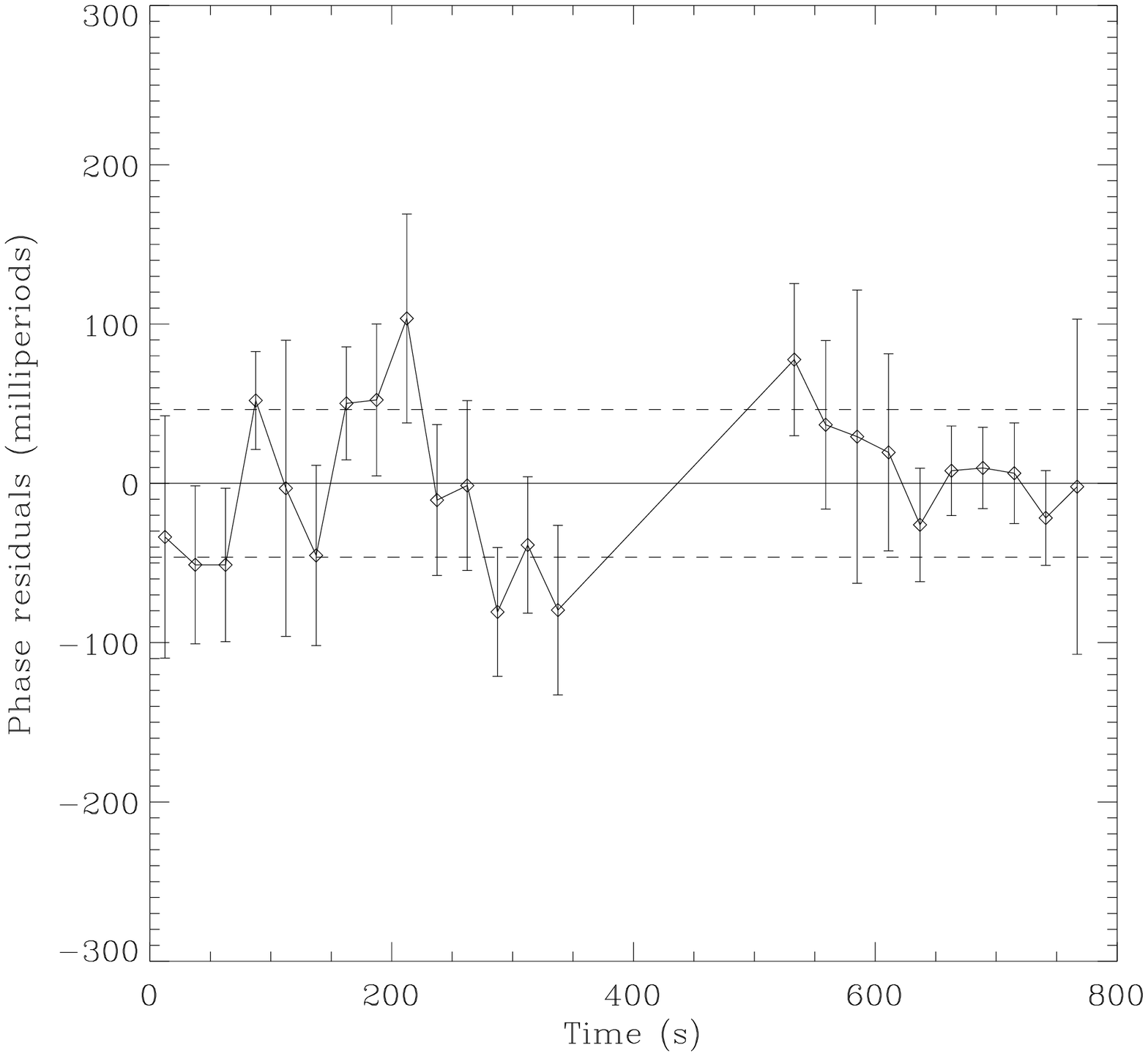]{Pulse phase residuals (observed - model) for our best
fitting orbit model.  The rms deviation is 
$\sim 46$ milliperiods and is denoted by the dashed horizontal lines. There is
some modest indication of an additional phase jitter in the first pulse 
train. See the caption to Figure 3 for the reference epoch. 
\label{fig7}} 

\figcaption[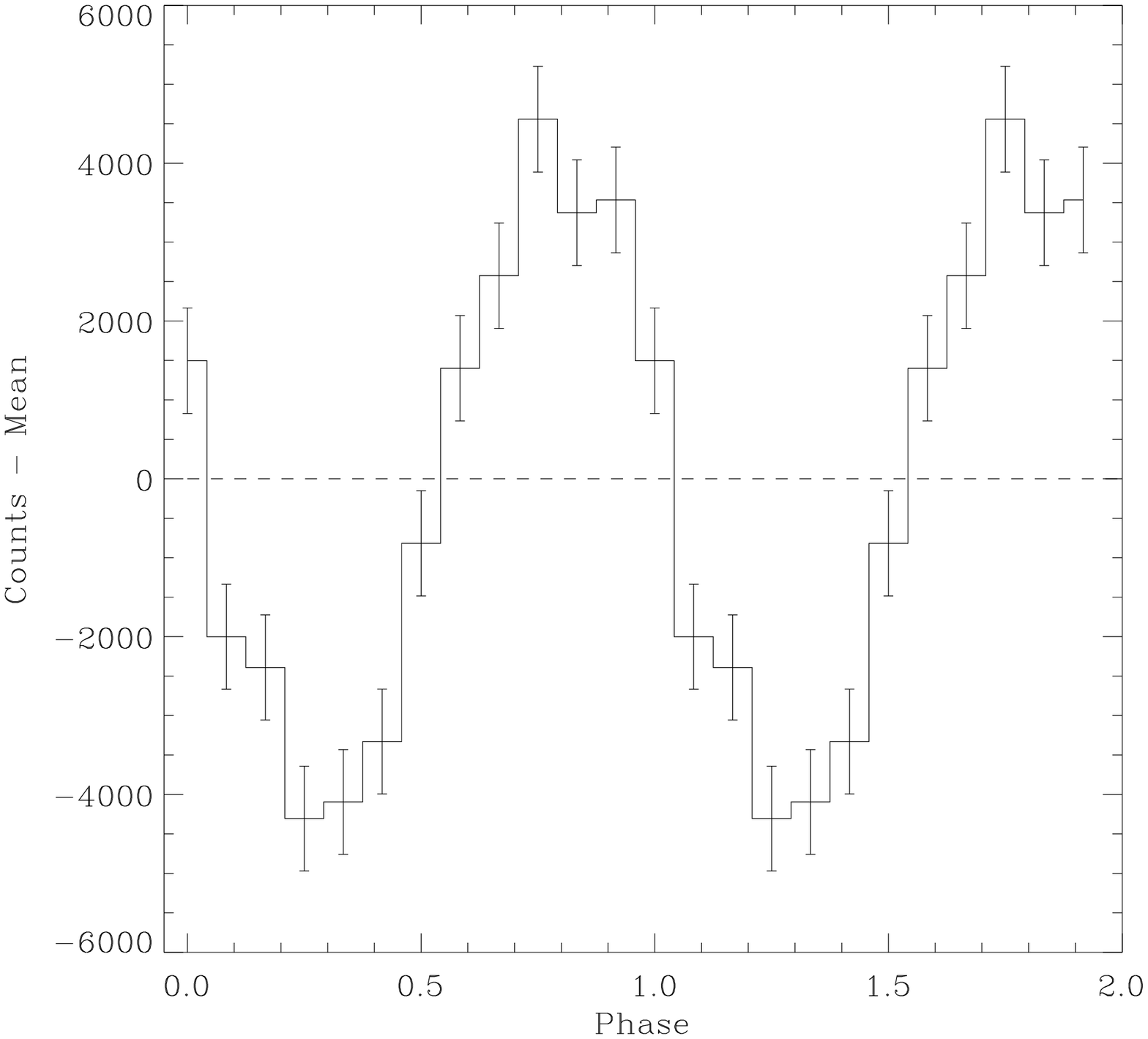]{Average pulse profile in the full 2 - 60 keV band 
produced by folding the two observed pulse trains with the best fitting orbit 
model.  The mean number of counts, 447340, has been subtracted. The profile is 
sinusoidal and has a half amplitude of $1\%$ of the mean number of counts. 
Phase zero is arbitrary.
\label{fig8}}

\figcaption[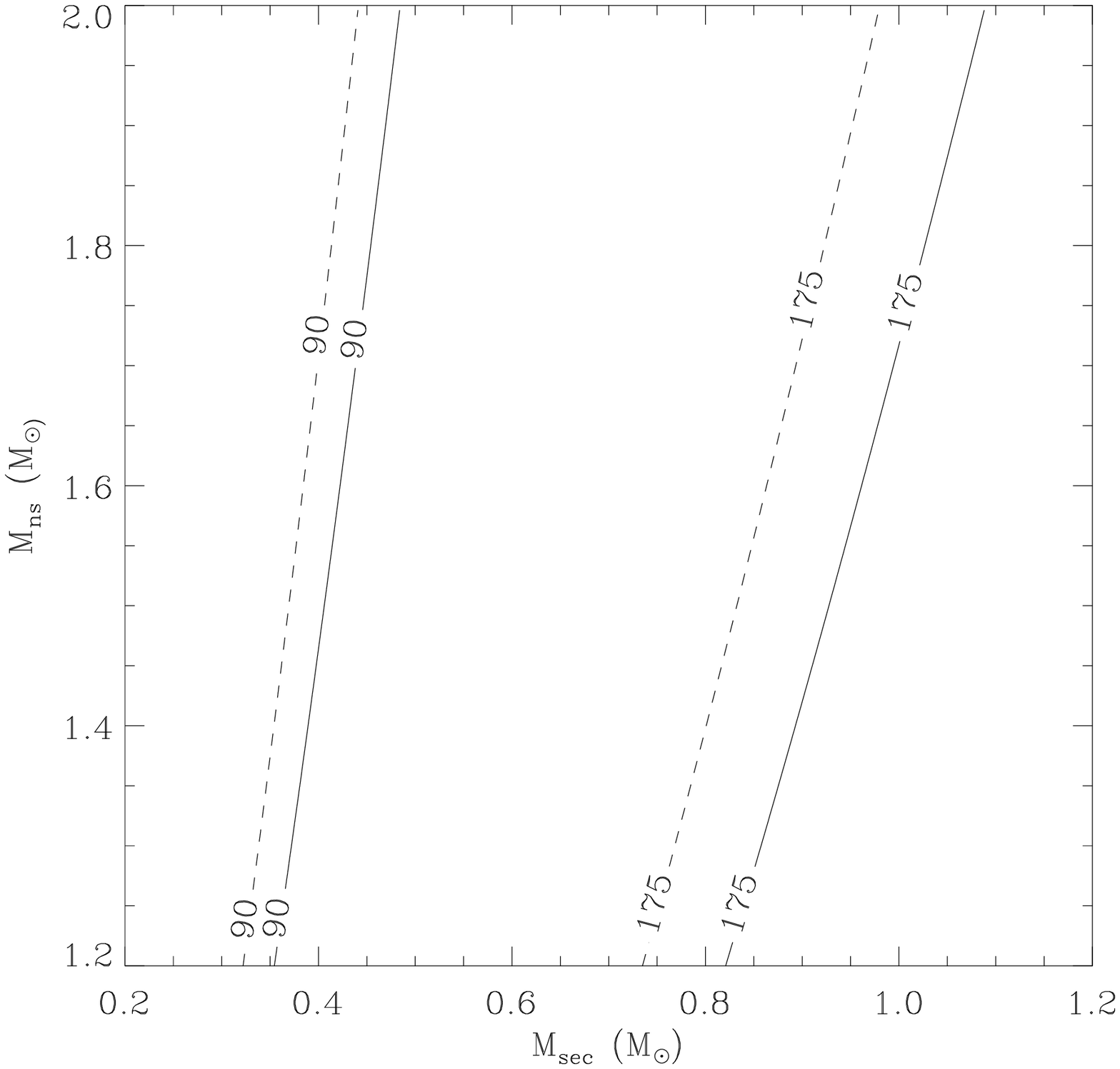]{Allowable component masses for 4U 1636-53 inferred from the 
neutron star orbital velocity constraints. A pair of velocity limits are 
plotted for two representative orbital inclinations, $60^{\circ}$ (solid), and 
$70^{\circ}$ (dashed). The curves are labelled with their respective
velocities.
 \label{fig9}}

\newpage

\begin{figure}
\begin{center}
 \includegraphics[width=6in, height=4in]{f1.ps}
\hspace{2in}%
\includegraphics[width=5in, height=4in]{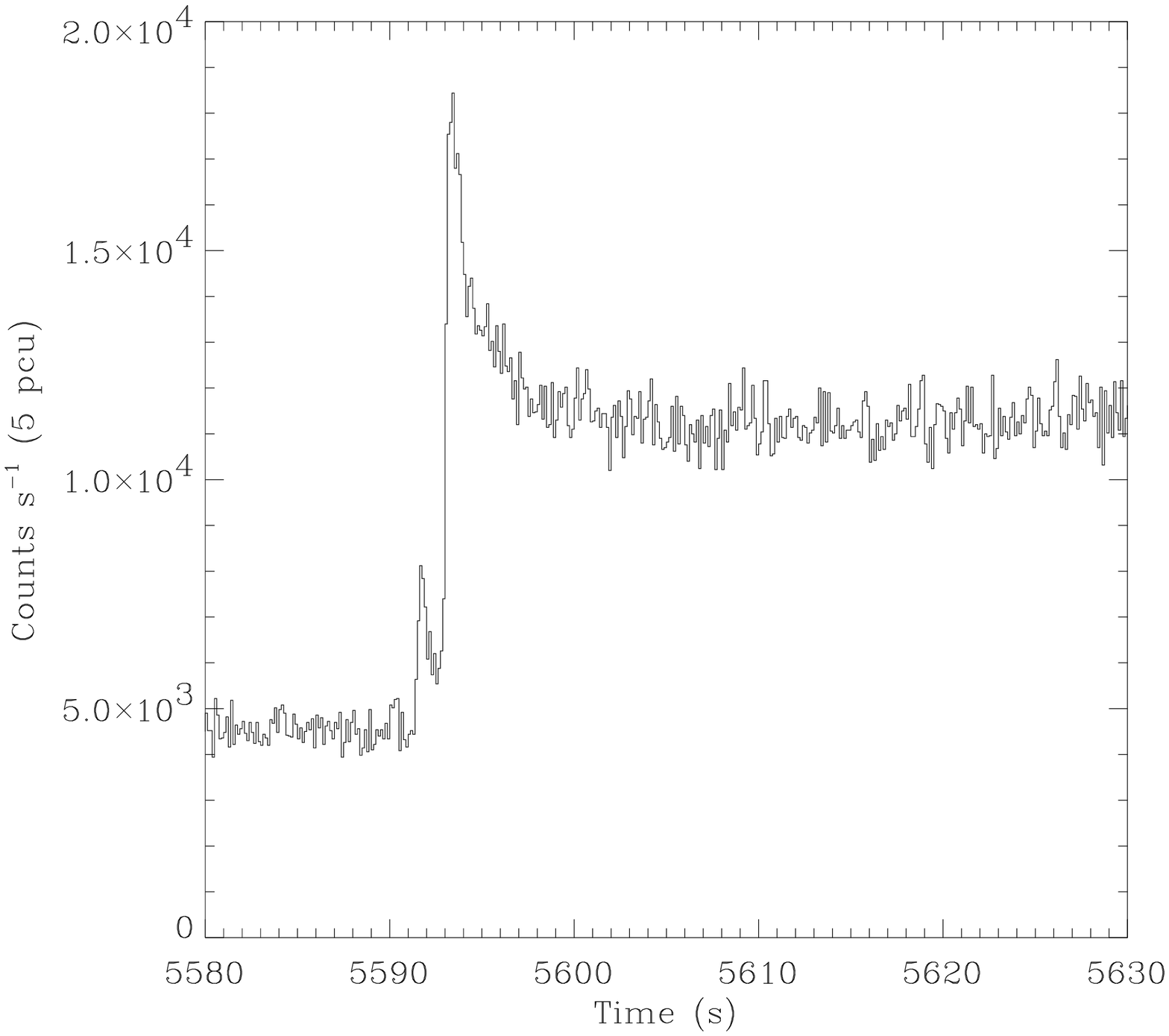}
\end{center}

Figure 1: Lightcurve of the superburst from 4U 1636-53 observed on
February 22, 2001 (top). The data are the 2 - 60 keV PCA countrates from the
Standard1 datamode. In the top panel the time resolution is 1 s. 
The dashed vertical lines denote the intervals with high time resolution event
mode data. The $\sim 800$ s interval in which pulsations 
are detected is shown by the vertical dotted lines. 
An exploded view of the sharp rise near the start of the burst
is also shown (bottom). Here the time resolution is 1/8 s. Note the double 
peaked profile which shows timescales typically seen in normal Type I X-ray 
bursts ($\sim 10$ s). Time is measured from 15:19:00 UTC on Feb. 22, 2001.
\end{figure}
\clearpage

\begin{figure}
\begin{center}
 \includegraphics[width=5in, height=4in]{f2.ps}
 \hspace{2in}%
\includegraphics[width=5in, height=4in]{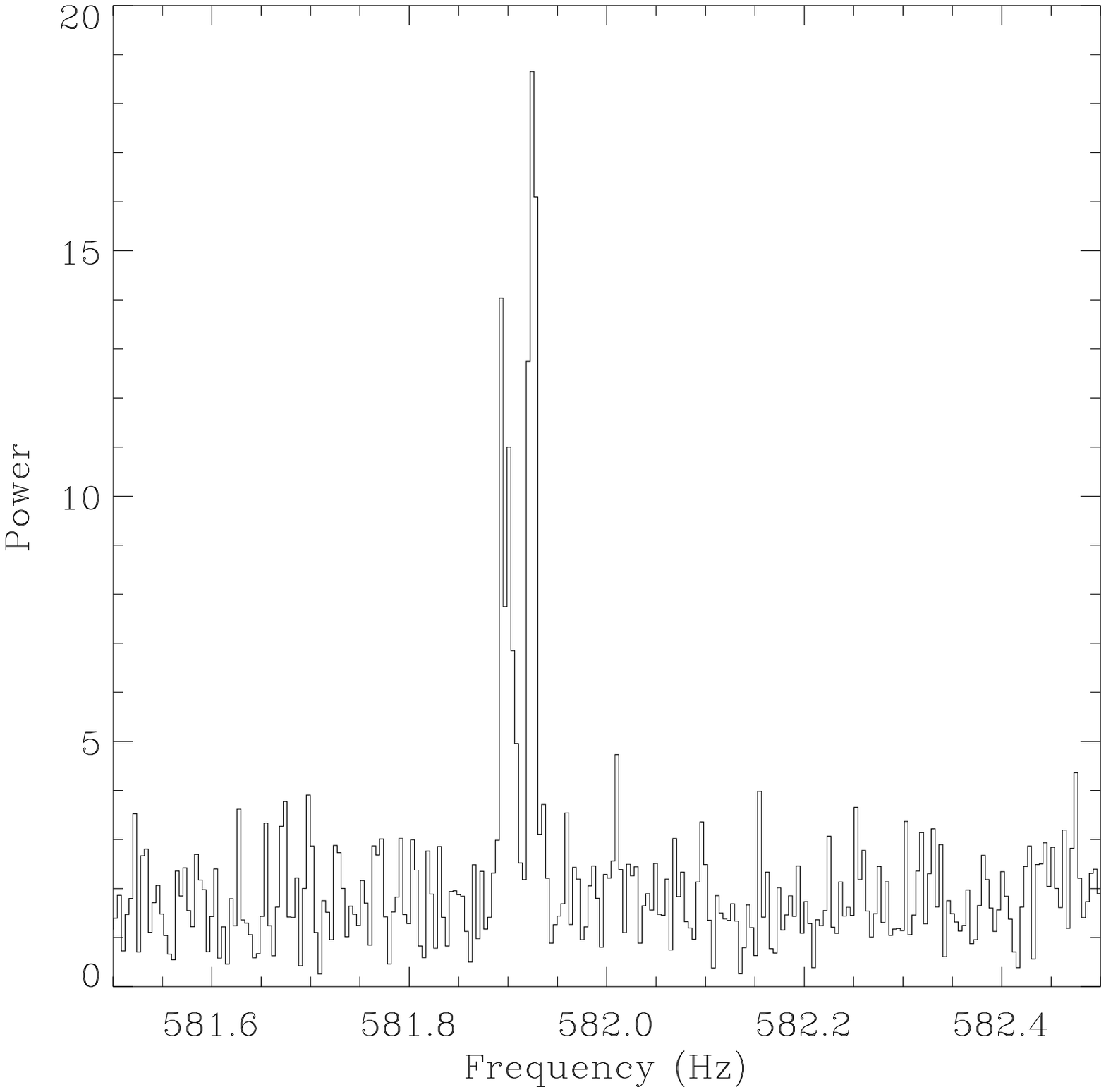}
\end{center}

Figure 2: Power spectrum (top) of the first 1024 second data interval 
starting at 17:00:44 UTC. The spectrum was computed using the full bandpass
event mode data. The Nyquist frequency is 4096 Hz. The frequency resolution 
is 1/128 Hz. The significant peak near 581 Hz is evident. An expanded view 
of the region around 582 Hz is also shown (bottom). The frequency resolution 
in this plot is 1/256 Hz. Two peaks separated by $\sim 0.03$ Hz are clearly 
resolved.
\end{figure}
\clearpage

\begin{figure}
\begin{center}
 \includegraphics[width=6in, height=6in]{f3.ps}
\end{center}
Figure 3: Dynamic $Z_1^2$ power spectrum of the 800 s interval 
containing significant pulsations. The $Z_1^2$ data were computed using
64 s data intervals and a new interval was started every 16 s. Contours
at 16, 18, 20, 22, 25, 30, 35, 40, 45, 50, 60, 70, 80, and 90 are shown. Two
pulse trains are clearly evident, separated by a gap of $\sim 200$ seconds.
The PCA countrate (4 detectors) is also shown (see right axes). Time is 
measured from $17:02:25$ UTC on Feb. 22, 2001.
\end{figure}
\clearpage

\begin{figure}
\begin{center}
 \includegraphics[width=6in, height=6in]{f4.ps}
\end{center}

Figure 4: Predicted orbital frequency evolution of a coherent 
pulsation centered on the neutron star in 4U 1636-53 and the orbital phase 
range of the 800 second pulsation interval during the superburst. We used the
optical ephemeris from Giles et al. (2002) along with the single assumption 
that optical maximum corresponds to superior conjunction of the optical
secondary. The vertical lines denote the phase range of the 
800 s pulsation interval.  The curve shows the frequency modulation which would
be produced with $v\sin i = 62$ km s$^{-1}$ for the neutron star. This is not 
intended to represent a fit to the observed frequency evolution, but 
rather to simply show the qualitative nature of the predicted frequency 
evolution.  Even with an offset of the phase relationship between optical 
max and superior conjunction of $\sim 0.1$ the ephemeris predicts that a 
pulsation frequency associated with the neutron star should be increasing.
\end{figure}
\clearpage

\begin{figure}
\begin{center}
 \includegraphics[width=6in, height=6in]{f5.ps}
\end{center}
Figure 5: Comparison of the FFT power spectrum which assumes no
frequency or phase evolution with the $Z_1^2$ power spectrum computed using
the best orbital frequency evolution model. The weak, broad peaks to the
left of the figure are from the standard FFT power spectrum, while the
sharp peaks to the right result from the phase coherent $Z_1^2$ power
spectrum. The sidelobe peaks in the coherent spectrum result from 
the gap between the two pulse trains.
\end{figure}
\clearpage

\begin{figure}
\begin{center}
 \includegraphics[width=6in, height=6in]{f6.ps}
\end{center}
Figure 6: Dynamic $Z_1^2$ contours as a function of frequency and
time. Also shown is the frequency evolution from the best orbit 
model. The solid curve shows the model with $\nu_0 = 582.1454$ Hz, 
$v_{ns}\sin i = 136$ km s$^{-1}$ and $\phi_{dyn} = 0.293$ and has a 
$Z_1^2 = 275.5$. See the caption to Figure 3 for the reference epoch. 
\end{figure}
\clearpage

\begin{figure}
\begin{center}
 \includegraphics[width=6in, height=6in]{f7.ps}
\end{center}
Figure 7: Pulse phase residuals (observed - model) for our best
fitting orbit model.  The rms deviation is 
$\sim 46$ milliperiods and is denoted by the dashed horizontal lines. There is
some modest indication of an additional phase jitter in the first pulse 
train. See the caption to Figure 3 for the reference epoch. 
\end{figure}
\clearpage

\begin{figure}
\begin{center}
 \includegraphics[width=6.5in, height=6in]{f8.ps}
\end{center}
Figure 8: Average pulse profile in the full 2 - 60 keV band 
produced by folding the two observed pulse trains with the best fitting orbit 
model.  The mean number of counts, 447340, has been subtracted. The profile is 
sinusoidal and has a half amplitude of $1\%$ of the mean number of counts. 
Phase zero is arbitrary.
\end{figure}
\clearpage

\begin{figure}
\begin{center}
 \includegraphics[width=6.0in, 
height=6in]{f9.ps}
\end{center}
\vskip 20pt
Figure 9: Allowable component masses for 4U 1636-53 inferred from the 
neutron star orbital velocity constraints. A pair of velocity limits are 
plotted for two representative orbital inclinations, $60^{\circ}$ (solid), and 
$70^{\circ}$ (dashed). The curves are labelled with their respective
velocities.

\end{figure}
\clearpage


\begin{references}
\reference{} Augusteijn, T., van der Hooft, F., de Jong, J.A., 
             van Kerkwijk, M.H., \& van Paradijs, J., 1998, A\&A, 332, 561
\reference{} Baraffe, I., Chabrier, G., Allard, F. \& Hauschildt, P. H. 1998, 
             A\&A, 337, 403
\reference{} Buccheri, R. et al. 1983, A\&A, 128, 245
\reference{} Chakrabarty, D. \& Morgan, E. H. 1998, Nature, 394, 346
\reference{} Cornelisse, R., Kuulkers, E., in't Zand, J. J.~M., 
             Verbunt, F. \& Heise, J.,  2002, A\&A, 382, 174
\reference{} Cornelisse, R., Heise, J., Kuulkers, E., Verbunt, F., \& 
             in't Zand, J. J.~M.  2000, \apj, 357, L21
\reference{} Cumming, A. \& Bildsten, L. 2000, ApJ, 544, 453
\reference{} Cumming, A. et al. 2001, ApJ, submitted, (astro-ph/0108009)
\reference{} Frank, J., King, A. R. \& Lasota, J.-P. 1987, A\&A, 178, 137
\reference{} Giles, A.B., Hill, K.M., Strohmayer, T. E. \& Cummings, N. 2002, 
             ApJ, 568, 222
\reference{} Heyl, J. S. 2001, astro-ph/0108450 
\reference{} in 't Zand, J. J. M. et al. 1998, A\&A, 331, L25
\reference{} in 't Zand, J.~J.~M.~et al.\ 2001, \aap, 372, 916
\reference{} Kolb, U., King, A. R. \& Baraffe, I. 2001, MNRAS, 321, 544
\reference{} Kuulkers, E. 2001, The Astronomers Telegram, \#68, 68, 1
\reference{} Kuulkers, E. et al. 2002, A\&A, 382, 503
\reference{} Markwardt, C. B. \& Swank, J. H. 2002, IAUC, 7870
\reference{} Miller, M.~C.\ 1999, \apj, 515, L77
\reference{} Morgan, E. H. \& Chakrabarty, D. 1998, Nature, 394, 346
\reference{} Muno, M. P. et al. 2002, ApJ, submitted (astro-ph/0204320)
\reference{} Muno, M.\ P., Fox, D.\ W., Morgan, E.\ H.\ \& Bildsten, L.\ 2000, 
             ApJ, 542, 1016
\reference{} Nash, S. G. 1984, SIAM J. Num. Anal., 21, 770
\reference{} Patterson, J. 1984, ApJS, 54, 443
\reference{} Remillard, R. A., Swank, J. H. \& Strohmayer, T. E. 2002, IAUC, 
             7893
\reference{} Smale, A.~P.~\& Mukai, K.\ 1988, \mnras, 231, 663 
\reference{} Spitkovsky, A., Levin, Y. \& Ushomirsky, G. 2001, ApJ, 566, 1018
\reference{} Standish, E. M., Newhall, X. X., Williams, J. G. \& Yeomans, D. K
             1992, in Explanatory Supplememt to the Astronomical Almanac, ed. 
             P. K. Seidelmann (Mill Valley: University Science), 239
\reference{} Strohmayer, T. E. \& Brown, E. F. 2002, ApJ, 566, 1045
\reference{} Strohmayer, T. E. 2001, Advances Sp. Res., vol. 28, Nos 2-3, 511
\reference{} Strohmayer, T.E., Zhang, W., \& Swank, J.H., 1997, ApJ, 487, L77
\reference{} Strohmayer, T. E., Zhang, W., Swank, J. H., White, N. E. \& 
             Lapidus, I. 1998a, ApJ, 498, L135
\reference{} Strohmayer, T.E., Zhang, W., Swank, J.H., \& Lapidus, I., 
             1998b, ApJ, 503, L147
\reference{} Strohmayer, T.E., \& Markwardt, C.B., 1999, ApJ, 516, L81
\reference{} Wijnands , R. 2001, \apjl, 554, L59
\reference{} Wijnands, R., \& van der Klis, M. 1998, Nature, 394, 344
\reference{} Zhang, W. W., Lapidus, I., Swank, J. H., White, N. E. \& 
             Titarchuk, L. 1997, IAUC, 6541

\end{references}
\end{document}